\documentclass[aps]{revtex4}
\usepackage{graphicx}
\usepackage{amssymb}
\usepackage{amsmath}
\usepackage{bm}
\usepackage{color}

\def\e{\varepsilon}

\def\p{\prime}
\def\s{\sigma}
\def\t{\theta}

\def\bk{{\bm k}}

\begin{document}

\title{Electric polarization in magnetic topological nodal semimetal thin films}
\author{Yuya Ominato$^{1}$, Ai Yamakage$^{2}$, and Kentaro Nomura$^{1,3}$}
\affiliation{Institute for Materials Research, Tohoku university, Sendai 980-8577, Japan}
\affiliation{Department of Physics, Nagoya University, Nagoya 464-8602, Japan}
\affiliation{Center for Spintronics Research Network, Tohoku University, Sendai 980-8577, Japan}
\date{\today}

\begin{abstract}
    We theoretically study the electric polarization in magnetic topological nodal semimetal thin films. In magnetically doped topological insulators, topological nodal semimetal phases emerge once the exchange coupling overcomes the band gap. Changing the magnetization direction, nodal structure is modulated and the system becomes topological nodal point or line semimetals. We find that nodal line semimetals are characterized by non-linear electric polarization, which is not observed in nodal point semimetals. The non-linear response originates from the existence of the surface states. Screening effect is self consistently included within a mean field approximation and the non-linear electric polarization is observed even in the presence of screening effect.
\end{abstract}

\maketitle

\section{Introduction}
\label{intro}

Topological semimetals are characterized by topological numbers and topological responses \cite{burkov2011weyl,burkov2011topological,ramamurthy2017quasitopological,armitage2018weyl}.
One of famous examples is a magnetic Weyl semimetal which is characterized by a Chern number defined on a closed surface enclosing a nodal point called Weyl point.
The Weyl points behave as a sink or source of Berry curvature, which leads to intrinsic anomalous Hall effect \cite{nagaosa2010anomalous}.
Magnetic Weyl semimetals exhibit a semiquantized anomalous Hall effect as a topological response \cite{burkov2011weyl,yang2011anomalous,burkov2014anomalous}. The intrinsic anomalous Hall effect in a magnetic Weyl semimetal is observed in recent experiments \cite{liu2018giant,wang2018large}.
On the other hand, nodal line semimetals are characterized by a Zak phase \cite{zak1989berry}. The Zak phase is closely related to the electric polarization \cite{smith1993theory,vanderbilt1993electric,resta1994macroscopic}, so that we expect non-trivial electric polarization in the nodal line semimetals \cite{hirayama2017topological,ramamurthy2017quasitopological}. However, the electric polarization is well defined only in insulators and the applicability of the theory of polarization based on the Zak phase is ambiguous in the nodal semimetals.

In this work, we theoretically study electric polarization in magnetic topological nodal semimetal thin films. We consider magnetically doped topological insulators \cite{zhang2013topology,chang2013experimental,chang2013thin,checkelsky2014trajectory,kou2014scale}. We assume ferromagnetic ordering of magnetic moments and introduce the exchange coupling between the band electrons and the magnetization. Once the exchange coupling overcomes the band gap, the system is brought into the magnetic topological semimetal phases \cite{burkov2011weyl,kurebayashi2014weyl}. The topological property of the band structure depends on the direction of the magnetization and the system becomes the nodal point and line semimetals by changing the direction of the magnetization \cite{burkov2011topological,habe2014three,burkov2018mirror}.
There are other candidate materials of the magnetic topological semimetals \cite{ueda2015magnetic,jin2017ferromagnetic}, where the topology of the electronic structure depends on the magnetic structure.
We numerically calculate the band structure of the topological semimetal thin films in the presence of the external potential. The screening effect is self-consistently included. We find that nodal line semimetals exhibit non-linear electric polarization, which originates from the existence of the surface states. In the nodal point semimetals, the non-linear electric polarization is not observed. Therefore, the nodal line semimetal phase is characterized by the non-linear electric polarization.

This work is organized as follows. In Section \ref{model_hamiltonian}, we introduce model Hamiltonian which describes electronic structures of topological insulators and exchange coupling between band electrons and local magnetic moments. The screened external potential is also added. We explain the band structure of bulk systems. In Section \ref{electronic_polarization}, we formulate a theoretical procedure to calculate a screened potential and a self-consistently determined electric polarization. In Section \ref{numerical_results}, we show numerically calculated band structures of the bulk and slab systems. We discuss the polarization of each state in the presence of the external potential
and the numerically calculated electric polarization. Conclusions are given in Section \ref{conclusions}.

\section{Model Hamiltonian}
\label{model_hamiltonian}

\begin{figure}
    \centering
    \includegraphics[width=10cm]{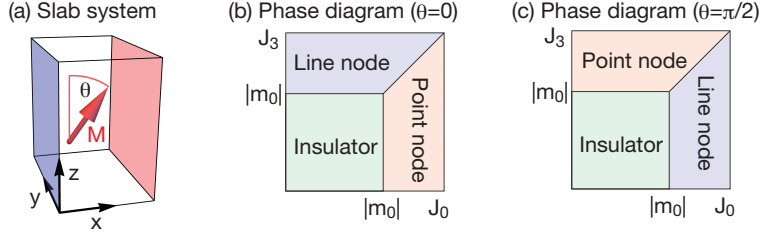}
    \caption{(a) Schematic picture of the slab system. $M$ represents magnetization set on the $x$-$z$ plane, and $\theta$ is an angle between the $z$ axis and $M$. We assume the open boundary condition for the $x$ direction, and the periodic boundary condition for the $y$ and $z$ directions. Topological phase diagram of bulk system for (b) $\theta=0$ and (c) $\theta=\pi/2$.}
    \label{fig_system}
\end{figure}

We start with the lattice model for three dimensional topological insulators (${\rm Bi}_2{\rm Se}_3$, ${\rm Bi}_2{\rm Te}_3$, and ${\rm Sb}_2{\rm Te}_3$) \cite{zhang2009topological,liu2010model},
\begin{align}
H_0=\tau_x\s_x t\sin{k_ya}-\tau_x \s_y t\sin{k_xa}+\tau_y t\sin{k_za}+m_\bk \tau_z,
\end{align}
where $t$ is a hopping parameter, $a$ is a lattice constant, and $m_\bk$ is a mass term,
\begin{align}
m_\bk&=m_0+m_2\sum_{i=x,y,z}(1-\cos k_i a).
\end{align}
$\bm{\s}$ and $\bm{\tau}$ are the Pauli matrices acting on the real spin and the pseudo spin (orbital) degrees of freedoms.
The exchange coupling between the band electrons and the local magnetic moments is written as \cite{wakatsuki2015domain}
\begin{align}
    H_{\rm ex}=J_0{\bm {\hat M}}\cdot{\bm \s}+J_3\tau_z{\bm {\hat M}}\cdot{\bm \s},
\end{align}
where $J_0, J_3$ are exchange coupling constants and ${\bm {\hat M}}$ is a unit vector representing the direction of magnetization. In this work, ${\bm {\hat M}}$ is set on the $x$-$z$ plane as we see in Figure.\ref{fig_system} (a). The first term is an usual exchange coupling and the second term originates from nonequality of the exchange coupling constants between two orbitals considered here, i.e., $p$-orbitals of (Bi,Sb) and (Te,Se). In magnetic topological insulators \cite{chang2013experimental,checkelsky2014trajectory,kou2014scale}, Cr atoms are substituted for Bi or Sb atoms. As a result, there is the nonequality of exchange coupling between $p$-orbitals of (Bi,Sb) and (Te,Se).
The total Hamiltonian is given as
\begin{align}
H_\bk=H_0+H_{\rm ex}+U(\bm{r}),
\end{align}
where $U(\bm{r})$ is the screened external potential. In the following calculation, we self-consistently include the screening effect and the detail of the numerical procedure is given in the next section.

The topological phase diagram of bulk system with no-external electric field is shown in Figure (\ref{fig_system}) (b) $\t=0$ and (c) $\t=\pi/2$. The system becomes topological semimetal phase when the exchange coupling overcomes the energy gap. In the present system, both of topological nodal line and point semimetal phases emerge. At $\t=0$, the system becomes a nodal line semimetal in $J_0<J_3$ and a nodal point semimetal in $J_0>J_3$. At $\t=\pi/2$, on the other hand, the system becomes a nodal line semimetal in $J_0>J_3$ and a nodal point semimetal in $J_0<J_3$. This means that the topological property of the electronic structure can be modulated by manipulating the magnetization direction.
The topologically non-trivial band structure in nodal line semimetals is characterized by the Zak phase
\begin{align}
    \t_{\rm Zak}(k_y,k_z)=-i\sum_{n}^{occ.}\int^{\pi/a}_{-\pi/a}\langle u_{n\bk}|\frac{\partial}{\partial k_x}|u_{n\bk}\rangle dk_x \hspace{3mm}({\rm mod} {~}2\pi),
\end{align}
where the summation is over the occupied states and $|u_{n\bk}\rangle$ is an eigenstate of the bulk Hamiltonian. In Section \ref{numerical_results}, we discuss the relation between the Zak phase and existence of surface states.

With a surface boundary for $x$ direction as represented in Figure \ref{fig_system} (a), the wave function is written as
\begin{align}
\psi(\bm{r})=\frac{1}{\sqrt{L_yL_z}}e^{i k_yy+i k_zz}\phi(x),
\end{align}
where $\psi(\bm{r})$ and $\phi(x)$ are four component spinors, and $L_y$, $L_z$ are the system sizes along $y$ and $z$ directions. The Schr\"{o}dinger equation becomes
\begin{align}
\e_{n\bk}\phi(x)=&\left[(m_0+3m_2)\tau_z+(J_0+J_3\tau_z)\hat{\bm{M}}\cdot\bm{\s}+U(x)\right]\phi(x) \notag \\
         &+\frac{it}{2}\tau_x\s_y\left[\phi(x+a)-\phi(x-a)\right]
          +t(\tau_x\s_x\sin k_ya+\tau_y\sin k_za)\phi(x) \notag \\
         &-\frac{m_2}{2}\tau_z\left[\phi(x+a)+\phi(x-a)\right] 
          -m_2\tau_z(\cos k_ya+\cos k_za)\phi(x).
\label{eq_sch}
\end{align}
We numerically diagonalize the above Hamiltonian and derive the eigenstates and the energy bands.

\section{Electric polarization induced by the screened potential}
\label{electronic_polarization}

We calculate electronic structure in the presence of the external electric field. The screening effect is self-consistently included within a mean field approximation by simultaneously solving the Schr\"{o}dinger equation and the Poisson's equation.
Our numerical procedure is as follows.
We set an initial value for the external potential as
\begin{align}
    U_0(x)=eE_0\left(x-\frac{N_xa}{2}\right),
\end{align}
where $N_xa$ is the width of the slab.
We derive eigenstates in the presence of the external potential.
We consider the slab as multilayers of zero-thickness planes with charge density $\rho^{\rm 2D}(x)$.
Using the derived eigenstates, $\rho^{\rm 2D}(x)$ is calculated as
\begin{align}
\rho^{\rm 2D}(x)=-e\sum_n\int_{\rm BZ}\frac{d^2k}{(2\pi)^2}f(\e_{n\bk})|\phi_{n\bk}(x)|^2,
\end{align}
where $f(\e_{n\bk})$ is the Fermi distribution function.
The electric field is calculated as
\begin{align}
E(x)=E_0+\frac{1}{2\kappa}\left(\sum_{j=0}^{n_x}\Delta\rho^{\rm 2D}(j a)-\sum_{j=n_x+1}^{N_x}\Delta\rho^{\rm 2D}(ja)\right),
\end{align}
where $\kappa$ is the permitivity and $n_x=\lfloor \frac{x}{a}\rfloor$.
$\Delta\rho^{2D}(ja)$ is the induced charge density, $\Delta\rho^{2D}(ja)=\rho^{2D}(ja)-\rho^{2D}_0(ja)$, where $\rho^{2D}_0(ja)$ is the charge density in the absence of the external potential.
Here, we assume charge neutrality of the slab, $E(x)=E_0\hspace{2mm}(x<0,L_x<x)$ or $\sum_j\Delta\rho^{2D}(ja)=0$.
Integrating the above electric field, the screened potential is calculated as
\begin{align}
U(x)=e\int_0^xE(x^\p)dx^\p+U(0),
\end{align}
where we set a boundary condition for the screened potential so that the average of the screened potential is zero, $\sum_j U(ja)$=0.
We perform numerical iteration so that the screened potential converges. The strength of the screening is characterized by a dimensionless parameter $\alpha$ defined as
\begin{align}
    \alpha=\frac{e^2}{\kappa ta}.
\end{align}
Using the eigenstates with the screened potential, we calculate the electric polarization of each eigenstate,
\begin{align}
    p_{n\bk}=-\frac{e}{L_yL_z}\sum_{j=0}^{N_x}|\phi_{n\bk}(ja)|^2\left(j-\frac{N_x}{2}\right).
\end{align}
The total electric polarization from the occupied eigenstates is written as
\begin{align}
    P_x&=\sum_{n\bk}f(\e_{n\bk})p_{n\bk} \notag \\
       &=-e\sum_n\int_{\rm BZ}\frac{d^2k}{(2\pi)^2}f(\e_{n\bk})\sum_{j=0}^{N_x}|\phi_{n\bk}(ja)|^2\left(j-\frac{N_x}{2}\right).
\end{align}
In the following section, we numerically diagonalize Equation (\ref{eq_sch}) and self-consistently determine the electric polarization.

\section{Numerical results}
\label{numerical_results}

\begin{figure}
    \centering
    \includegraphics[width=14cm]{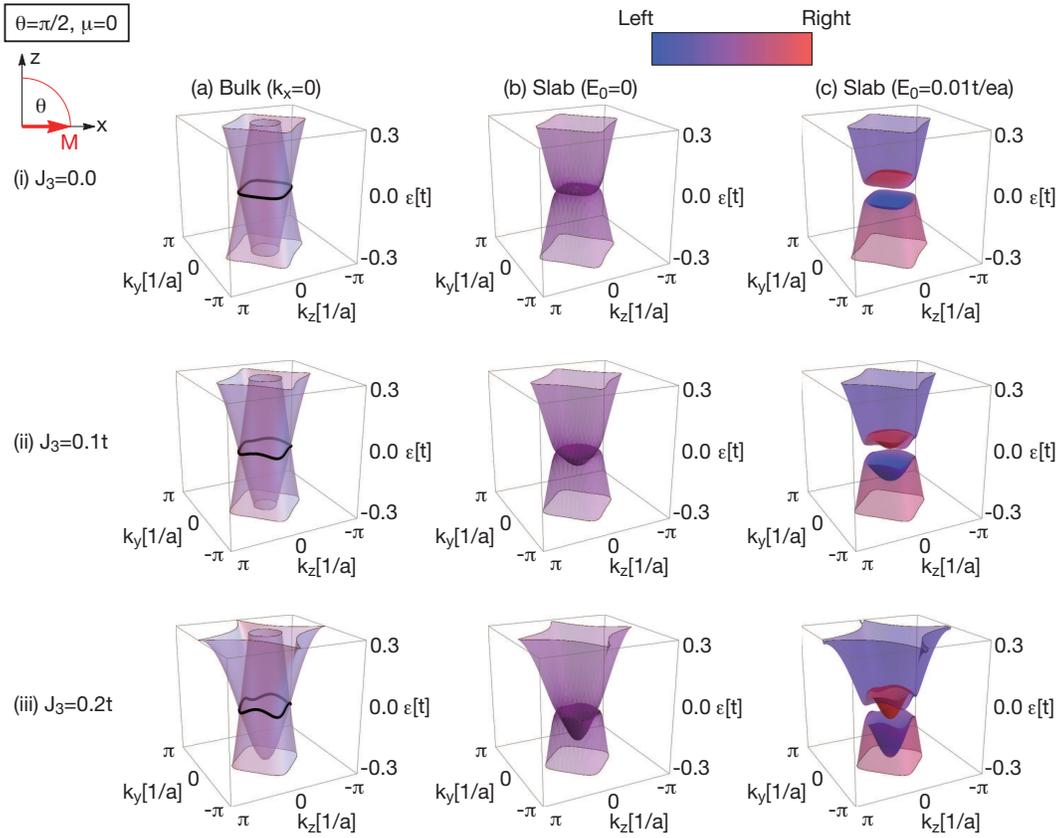}
    \caption{Band structure of nodal line semimetals (a) in the bulk system, (b) in the slab system with no E-field, and (c) in the slab system with E-field. We set the parameters as $m_2/t=1, m_0/t=-0.5, J_0/t=1, N_x=30,$ and $\alpha=0.1$.}
    \label{fig_nls_band}
\end{figure}

\begin{figure}
    \centering
    \includegraphics[width=10cm]{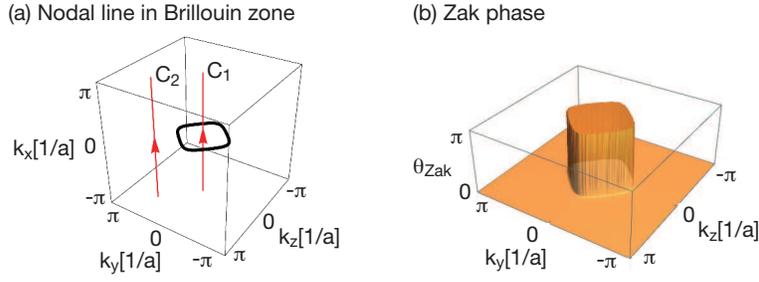}
    \caption{(a) The black closed curve represents nodal line in the Brillouin zone. Two paths $C_1$ and $C_2$ are depicted by red arrows and give the Zak phase $\pi$ and $0$, respectively. (b) The Zak phase calculated at fixed ($k_y,k_z$). The Zak phase is quantized as $\pi$ in the area enclosed by the nodal line projectoed on $k_y$-$k_z$ plane and it is zero in the other area.}
    \label{fig_zak}
\end{figure}

As represented in Figure \ref{fig_system} (a), we consider the open boundary condition for the $x$ direction, and the periodic boundary condition for the $y$ and $z$ directions.
We set the parameters as $m_2/t=1, m_0/t=-0.5, J_0/t=1$, and $N_x=30$. First, we consider a nodal line semimetal phase, $\t=\pi/2$. Figure \ref{fig_nls_band} shows the band structure of nodal line semimetals for several parameters.
At $J_3=0$, there is a nodal line on the $k_x=0$ plane in the bulk system.
In Figure \ref{fig_zak}, we see that the Zak phase is quantized as $\pi$ when the integration path go through the area enclosed by the nodal line. Otherwise, the Zak phase becomes zero. In the slab system, the wave number $k_x$ is discretized and we obtain subband structures. In Figure \ref{fig_nls_band} (b), we show the lowest subbands. There are degenerate flat bands in the wave numbers $(k_y,k_z)$ enclosed by the nodal line where the Zak phase is quantized as $\pi$. The flat band states are localized on the left and right surfaces. In the presence of the external potential, the degenerate flat bands are gapped as shown in Figure \ref{fig_nls_band} (c). The colored bar represents the polarization of each state.
We see that the surface states localized on the left surface are occupied and on the right surface are empty. It means that the surface states are completely polarized and the surface electric polarization occurs. The occupied bulk states are polarized to the opposite direction so that the surface electric polarization is screened. At finite $J_3$, the qualitative behavior is the same as $J_3=0$ but the band structure is slightly modified. The nodal line is not bounded on $\e=0$ and the surface bands are bended. As a result, a larger external potential is required in order to achieve complete surface polarization compared with the flat surface bands i.e., $J_3=0$.

\begin{figure}
    \centering
    \includegraphics[width=14cm]{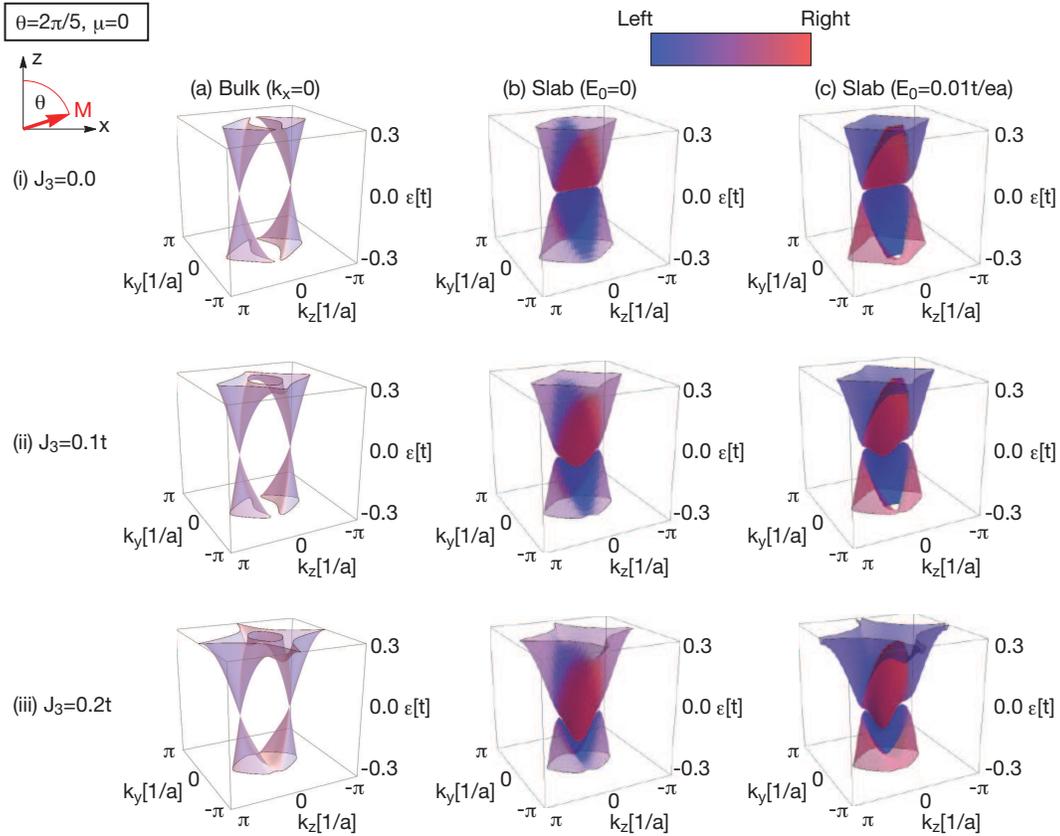}
    \caption{Band structure of nodal point (Weyl) semimetals (a) in the bulk system, (b) in the slab system with no external potential, and (c) in the slab system with an external potential. We set the parameters as $m_2/t=1, m_0/t=-0.5, J_0/t=1, N_x=30,$ and $ \alpha=0.1$.}
    \label{fig_nps_band}
\end{figure}

Figure \ref{fig_nps_band} shows the band structure of nodal point (Weyl) semimetal phase, $\t=2\pi/5$. When the magnetization has a finite component along $z$ direction, the nodal line is gapped except for a pair of nodal points and the system becomes nodal point (Weyl) semimetals. In the slab system, there are chiral surface states localized on the left or right surfaces and they are seamlessly connected to the bulk states. The velocity of the chiral surface states increases with the decrease of $\t$ and it becomes maximum at $\t=0$. In the presence of the external potential, the left surface states are pushed down and the right surface states are pushed up. Consequently, the surface electric polarization occurs but it is weak compared with that in the nodal line semimetals. In the nodal line semimetals, we observe the surface electric polarization which is proportional to the number of surface states in the sufficiently strong external potential. On the other hand, the required external potential to achieve the surface electric polarization which is proportional to the number of the surface states in the nodal point semimetals is much larger than that in the nodal line semimetals.

\begin{figure}
    \centering
    \includegraphics[width=13cm]{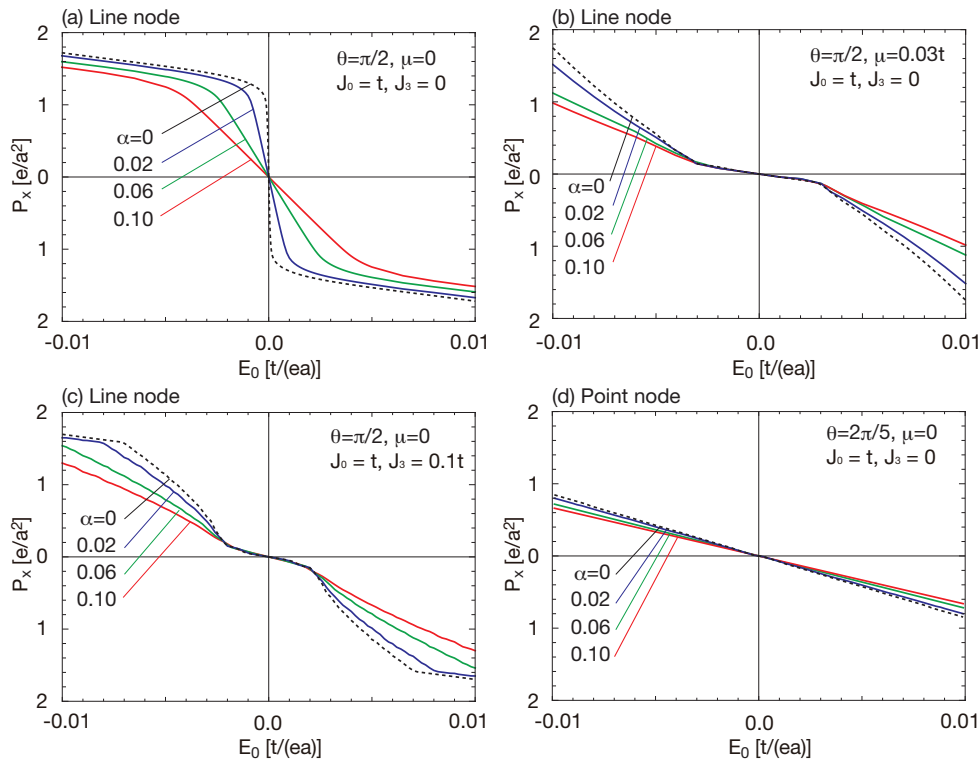}
    \caption{Polarization as a function of the external electric field for several screening strength $\alpha$.}
    \label{fig_polarization_e}
\end{figure}

Using the self-consistently derived electronic states, we calculate the electric polarization in the slab system. Figure \ref{fig_polarization_e} shows the electric polarization as a function of the external electric field for several parameters. In the nodal line semimetal, Figure \ref{fig_polarization_e} (a), we see that the system exhibits non-linear electric polarization.
In the absence of the screening effect $(\alpha=0)$, the electric polarization abruptly changes at $E_0=0$.
In the bulk limit, the electric polarization is induced by infinitesimal electric field and discontinuously changes at $E_0=0$. However, in our numerical calculation, the electric polarization continuously changes at $E_0=0$ because of the finite size effect.
In the presence of the screening effect, the step structure is broadened and the electric polarization vanishes at $E_0=0$. However, the kink structure of the electric polarization is preserved. Figure \ref{fig_polarization_e} (b) shows the electric polarization for finite chemical potential $\mu/t=0.03$. Even in the finite chemical potential, the kink structure is observed, though the kink appears at a finite electric field. This kink structure is broadened by the screening effect.
In finite $J_3$ term, Figure \ref{fig_polarization_e} (c), the surface bands are bended and the situation is similar to the finite chemical potential case, Figure \ref{fig_polarization_e} (b), so that the qualitative behavior is the same. In the nodal point semimetal, Figure \ref{fig_polarization_e} (d), the electric polarization is proportional to the external electric field and there is no kink structure. Therefore, the kink structure is characteristic to the nodal line semimetals and is closely related to the topological property of the system.

\begin{figure}
    \centering
    \includegraphics[width=14cm]{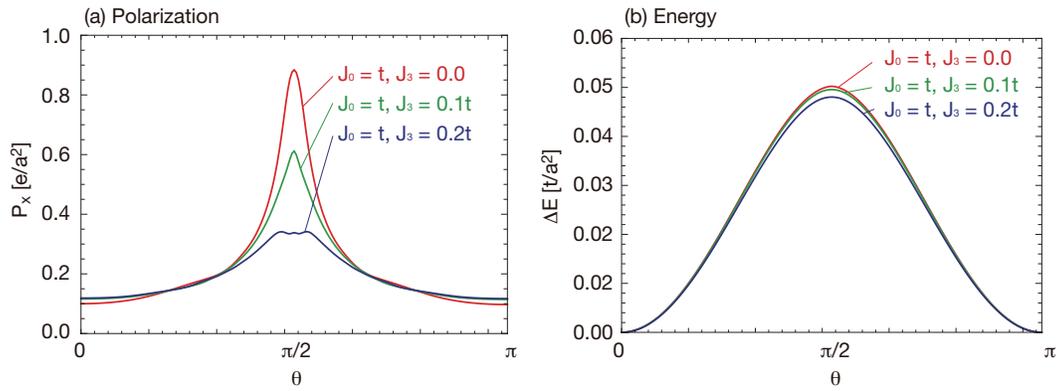}
    \caption{(a) The total polarization and (b) the total energy as a function of the angle $\t$. We set $E_0=0.01t/(ea)$, $\alpha=0.1$, and $\mu=0$.}
    \label{fig_polarization_theta}
\end{figure}

Figure \ref{fig_polarization_theta} (a) shows the electric polarization as a function of the angle $\t$. There are peak structure of the electric polarization at $\t=\pi/2$ because the system becomes the nodal line semimetal at this angle and exhibits large electric polarization originating from the surface states. With the increase of $J_3$, the peak value becomes smaller. This is because the surface bands are bended and the kink structure is broadened. Figure \ref{fig_polarization_theta} (b) shows variation of the total energy defined as
\begin{align}
    \Delta E(\t)=\frac{1}{N_x}\int_{\rm BZ}\frac{d^2k}{(2\pi)^2}\sum_{n}\left[\e_{n\bk}(\t)f\left(\e_{n\bk}(\t)\right)-\e_{n\bk}(\t=0)f\left(\e_{n\bk}(\t=0)\right)\right],
\end{align}
where $\e_{n\bk}(\t)$ is an eigenenergy of the slab system with the magnetization angle $\t$.
The variation of the total energy also becomes maximum at $\t=\pi/2$ and decreases away from $\t=\pi/2$. Unlike the electric polarization, $\Delta\e(\t)$ weakly depends on $J_3$ and varies slowly compared with the electric polarization. This is because the electric polarization is almost determined by the property of the surface states but the variation of the total energy is almost determined by the bulk band structure. Figure \ref{fig_polarization_theta} (b) means that the magnetization easy axis is parallel to the surface.

\section{Conclusions}
\label{conclusions}

We investigated the electronic structure in the magnetically doped topological insulator thin films. We found that the system becomes topological nodal semimetal phases when the exchange coupling exceeds the band gap and the topological property of the band structure is modulated by changing the magnetization direction. We calculated the electric polarization in the topological nodal semimetal thin films. We self-consistently include the screening effect and find that the nodal line semimetals exhibit non-linear electric polarization even in a finite chemical potential. This is a topological response and the nodal line semimetals are characterized by the non-linear electric polarization, which is not observed in the nodal point (Weyl) semimetals.

\section*{ACKNOWLEDGMENT}
This work was supported by Kakenhi Grants-in-Aid (Nos. JP15H05851,JP15H05854 and JP17K05485) from the Japan Society for the Promotion of Science (JSPS) and JST CREST Grant Number JPMJCR18T2.

\bibliography{nodal_semimetal_polarization}

\begin{thebibliography}{27}
\expandafter\ifx\csname natexlab\endcsname\relax\def\natexlab#1{#1}\fi
\expandafter\ifx\csname bibnamefont\endcsname\relax
  \def\bibnamefont#1{#1}\fi
\expandafter\ifx\csname bibfnamefont\endcsname\relax
  \def\bibfnamefont#1{#1}\fi
\expandafter\ifx\csname citenamefont\endcsname\relax
  \def\citenamefont#1{#1}\fi
\expandafter\ifx\csname url\endcsname\relax
  \def\url#1{\texttt{#1}}\fi
\expandafter\ifx\csname urlprefix\endcsname\relax\def\urlprefix{URL }\fi
\providecommand{\bibinfo}[2]{#2}
\providecommand{\eprint}[2][]{\url{#2}}

\bibitem[{\citenamefont{Burkov and Balents}(2011)}]{burkov2011weyl}
\bibinfo{author}{\bibfnamefont{A.}~\bibnamefont{Burkov}} \bibnamefont{and}
  \bibinfo{author}{\bibfnamefont{L.}~\bibnamefont{Balents}},
  \bibinfo{journal}{Physical review letters} \textbf{\bibinfo{volume}{107}},
  \bibinfo{pages}{127205} (\bibinfo{year}{2011}).

\bibitem[{\citenamefont{Burkov et~al.}(2011)\citenamefont{Burkov, Hook, and
  Balents}}]{burkov2011topological}
\bibinfo{author}{\bibfnamefont{A.}~\bibnamefont{Burkov}},
  \bibinfo{author}{\bibfnamefont{M.}~\bibnamefont{Hook}}, \bibnamefont{and}
  \bibinfo{author}{\bibfnamefont{L.}~\bibnamefont{Balents}},
  \bibinfo{journal}{Physical Review B} \textbf{\bibinfo{volume}{84}},
  \bibinfo{pages}{235126} (\bibinfo{year}{2011}).

\bibitem[{\citenamefont{Ramamurthy and
  Hughes}(2017)}]{ramamurthy2017quasitopological}
\bibinfo{author}{\bibfnamefont{S.~T.} \bibnamefont{Ramamurthy}}
  \bibnamefont{and} \bibinfo{author}{\bibfnamefont{T.~L.}
  \bibnamefont{Hughes}}, \bibinfo{journal}{Physical Review B}
  \textbf{\bibinfo{volume}{95}}, \bibinfo{pages}{075138}
  (\bibinfo{year}{2017}).

\bibitem[{\citenamefont{Armitage et~al.}(2018)\citenamefont{Armitage, Mele, and
  Vishwanath}}]{armitage2018weyl}
\bibinfo{author}{\bibfnamefont{N.}~\bibnamefont{Armitage}},
  \bibinfo{author}{\bibfnamefont{E.}~\bibnamefont{Mele}}, \bibnamefont{and}
  \bibinfo{author}{\bibfnamefont{A.}~\bibnamefont{Vishwanath}},
  \bibinfo{journal}{Reviews of Modern Physics} \textbf{\bibinfo{volume}{90}},
  \bibinfo{pages}{015001} (\bibinfo{year}{2018}).

\bibitem[{\citenamefont{Nagaosa et~al.}(2010)\citenamefont{Nagaosa, Sinova,
  Onoda, MacDonald, and Ong}}]{nagaosa2010anomalous}
\bibinfo{author}{\bibfnamefont{N.}~\bibnamefont{Nagaosa}},
  \bibinfo{author}{\bibfnamefont{J.}~\bibnamefont{Sinova}},
  \bibinfo{author}{\bibfnamefont{S.}~\bibnamefont{Onoda}},
  \bibinfo{author}{\bibfnamefont{A.}~\bibnamefont{MacDonald}},
  \bibnamefont{and} \bibinfo{author}{\bibfnamefont{N.}~\bibnamefont{Ong}},
  \bibinfo{journal}{Reviews of modern physics} \textbf{\bibinfo{volume}{82}},
  \bibinfo{pages}{1539} (\bibinfo{year}{2010}).

\bibitem[{\citenamefont{Yang et~al.}(2011)\citenamefont{Yang, Lu, and
  Ran}}]{yang2011anomalous}
\bibinfo{author}{\bibfnamefont{K.-Y.} \bibnamefont{Yang}},
  \bibinfo{author}{\bibfnamefont{Y.-M.} \bibnamefont{Lu}}, \bibnamefont{and}
  \bibinfo{author}{\bibfnamefont{Y.}~\bibnamefont{Ran}},
  \bibinfo{journal}{Phys. Rev. B} \textbf{\bibinfo{volume}{84}},
  \bibinfo{pages}{075129} (\bibinfo{year}{2011}).

\bibitem[{\citenamefont{Burkov}(2014)}]{burkov2014anomalous}
\bibinfo{author}{\bibfnamefont{A.~A.} \bibnamefont{Burkov}},
  \bibinfo{journal}{Phys. Rev. Lett.} \textbf{\bibinfo{volume}{113}},
  \bibinfo{pages}{187202} (\bibinfo{year}{2014}).

\bibitem[{\citenamefont{Liu et~al.}(2018)\citenamefont{Liu, Sun, Kumar,
  Muechler, Sun, Jiao, Yang, Liu, Liang, Xu et~al.}}]{liu2018giant}
\bibinfo{author}{\bibfnamefont{E.}~\bibnamefont{Liu}},
  \bibinfo{author}{\bibfnamefont{Y.}~\bibnamefont{Sun}},
  \bibinfo{author}{\bibfnamefont{N.}~\bibnamefont{Kumar}},
  \bibinfo{author}{\bibfnamefont{L.}~\bibnamefont{Muechler}},
  \bibinfo{author}{\bibfnamefont{A.}~\bibnamefont{Sun}},
  \bibinfo{author}{\bibfnamefont{L.}~\bibnamefont{Jiao}},
  \bibinfo{author}{\bibfnamefont{S.-Y.} \bibnamefont{Yang}},
  \bibinfo{author}{\bibfnamefont{D.}~\bibnamefont{Liu}},
  \bibinfo{author}{\bibfnamefont{A.}~\bibnamefont{Liang}},
  \bibinfo{author}{\bibfnamefont{Q.}~\bibnamefont{Xu}}, \bibnamefont{et~al.},
  \bibinfo{journal}{Nature Physics}  (\bibinfo{year}{2018}), ISSN
  \bibinfo{issn}{1745-2481},
  \urlprefix\url{https://doi.org/10.1038/s41567-018-0234-5}.

\bibitem[{\citenamefont{Wang et~al.}(2018)\citenamefont{Wang, Xu, Lou, Liu, Li,
  Huang, Shen, Weng, Wang, and Lei}}]{wang2018large}
\bibinfo{author}{\bibfnamefont{Q.}~\bibnamefont{Wang}},
  \bibinfo{author}{\bibfnamefont{Y.}~\bibnamefont{Xu}},
  \bibinfo{author}{\bibfnamefont{R.}~\bibnamefont{Lou}},
  \bibinfo{author}{\bibfnamefont{Z.}~\bibnamefont{Liu}},
  \bibinfo{author}{\bibfnamefont{M.}~\bibnamefont{Li}},
  \bibinfo{author}{\bibfnamefont{Y.}~\bibnamefont{Huang}},
  \bibinfo{author}{\bibfnamefont{D.}~\bibnamefont{Shen}},
  \bibinfo{author}{\bibfnamefont{H.}~\bibnamefont{Weng}},
  \bibinfo{author}{\bibfnamefont{S.}~\bibnamefont{Wang}}, \bibnamefont{and}
  \bibinfo{author}{\bibfnamefont{H.}~\bibnamefont{Lei}},
  \bibinfo{journal}{Nature Communications} \textbf{\bibinfo{volume}{9}},
  \bibinfo{pages}{3681} (\bibinfo{year}{2018}), ISSN \bibinfo{issn}{2041-1723},
  \urlprefix\url{https://doi.org/10.1038/s41467-018-06088-2}.

\bibitem[{\citenamefont{Zak}(1989)}]{zak1989berry}
\bibinfo{author}{\bibfnamefont{J.}~\bibnamefont{Zak}},
  \bibinfo{journal}{Physical review letters} \textbf{\bibinfo{volume}{62}},
  \bibinfo{pages}{2747} (\bibinfo{year}{1989}).

\bibitem[{\citenamefont{King-Smith and Vanderbilt}(1993)}]{smith1993theory}
\bibinfo{author}{\bibfnamefont{R.~D.} \bibnamefont{King-Smith}}
  \bibnamefont{and}
  \bibinfo{author}{\bibfnamefont{D.}~\bibnamefont{Vanderbilt}},
  \bibinfo{journal}{Phys. Rev. B} \textbf{\bibinfo{volume}{47}},
  \bibinfo{pages}{1651} (\bibinfo{year}{1993}),
  \urlprefix\url{https://link.aps.org/doi/10.1103/PhysRevB.47.1651}.

\bibitem[{\citenamefont{Vanderbilt and
  King-Smith}(1993)}]{vanderbilt1993electric}
\bibinfo{author}{\bibfnamefont{D.}~\bibnamefont{Vanderbilt}} \bibnamefont{and}
  \bibinfo{author}{\bibfnamefont{R.~D.} \bibnamefont{King-Smith}},
  \bibinfo{journal}{Phys. Rev. B} \textbf{\bibinfo{volume}{48}},
  \bibinfo{pages}{4442} (\bibinfo{year}{1993}),
  \urlprefix\url{https://link.aps.org/doi/10.1103/PhysRevB.48.4442}.

\bibitem[{\citenamefont{Resta}(1994)}]{resta1994macroscopic}
\bibinfo{author}{\bibfnamefont{R.}~\bibnamefont{Resta}},
  \bibinfo{journal}{Reviews of modern physics} \textbf{\bibinfo{volume}{66}},
  \bibinfo{pages}{899} (\bibinfo{year}{1994}).

\bibitem[{\citenamefont{Hirayama et~al.}(2017)\citenamefont{Hirayama, Okugawa,
  Miyake, and Murakami}}]{hirayama2017topological}
\bibinfo{author}{\bibfnamefont{M.}~\bibnamefont{Hirayama}},
  \bibinfo{author}{\bibfnamefont{R.}~\bibnamefont{Okugawa}},
  \bibinfo{author}{\bibfnamefont{T.}~\bibnamefont{Miyake}}, \bibnamefont{and}
  \bibinfo{author}{\bibfnamefont{S.}~\bibnamefont{Murakami}},
  \bibinfo{journal}{Nature communications} \textbf{\bibinfo{volume}{8}},
  \bibinfo{pages}{14022} (\bibinfo{year}{2017}).

\bibitem[{\citenamefont{Zhang et~al.}(2013)\citenamefont{Zhang, Chang, Tang,
  Zhang, Feng, Li, Wang, Chen, Liu, Duan et~al.}}]{zhang2013topology}
\bibinfo{author}{\bibfnamefont{J.}~\bibnamefont{Zhang}},
  \bibinfo{author}{\bibfnamefont{C.-Z.} \bibnamefont{Chang}},
  \bibinfo{author}{\bibfnamefont{P.}~\bibnamefont{Tang}},
  \bibinfo{author}{\bibfnamefont{Z.}~\bibnamefont{Zhang}},
  \bibinfo{author}{\bibfnamefont{X.}~\bibnamefont{Feng}},
  \bibinfo{author}{\bibfnamefont{K.}~\bibnamefont{Li}},
  \bibinfo{author}{\bibfnamefont{L.-l.} \bibnamefont{Wang}},
  \bibinfo{author}{\bibfnamefont{X.}~\bibnamefont{Chen}},
  \bibinfo{author}{\bibfnamefont{C.}~\bibnamefont{Liu}},
  \bibinfo{author}{\bibfnamefont{W.}~\bibnamefont{Duan}}, \bibnamefont{et~al.},
  \bibinfo{journal}{Science} \textbf{\bibinfo{volume}{339}},
  \bibinfo{pages}{1582} (\bibinfo{year}{2013}).

\bibitem[{\citenamefont{Chang et~al.}(2013{\natexlab{a}})\citenamefont{Chang,
  Zhang, Feng, Shen, Zhang, Guo, Li, Ou, Wei, Wang
  et~al.}}]{chang2013experimental}
\bibinfo{author}{\bibfnamefont{C.-Z.} \bibnamefont{Chang}},
  \bibinfo{author}{\bibfnamefont{J.}~\bibnamefont{Zhang}},
  \bibinfo{author}{\bibfnamefont{X.}~\bibnamefont{Feng}},
  \bibinfo{author}{\bibfnamefont{J.}~\bibnamefont{Shen}},
  \bibinfo{author}{\bibfnamefont{Z.}~\bibnamefont{Zhang}},
  \bibinfo{author}{\bibfnamefont{M.}~\bibnamefont{Guo}},
  \bibinfo{author}{\bibfnamefont{K.}~\bibnamefont{Li}},
  \bibinfo{author}{\bibfnamefont{Y.}~\bibnamefont{Ou}},
  \bibinfo{author}{\bibfnamefont{P.}~\bibnamefont{Wei}},
  \bibinfo{author}{\bibfnamefont{L.-L.} \bibnamefont{Wang}},
  \bibnamefont{et~al.}, \bibinfo{journal}{Science}
  \textbf{\bibinfo{volume}{340}}, \bibinfo{pages}{167}
  (\bibinfo{year}{2013}{\natexlab{a}}).

\bibitem[{\citenamefont{Chang et~al.}(2013{\natexlab{b}})\citenamefont{Chang,
  Zhang, Liu, Zhang, Feng, Li, Wang, Chen, Dai, Fang et~al.}}]{chang2013thin}
\bibinfo{author}{\bibfnamefont{C.-Z.} \bibnamefont{Chang}},
  \bibinfo{author}{\bibfnamefont{J.}~\bibnamefont{Zhang}},
  \bibinfo{author}{\bibfnamefont{M.}~\bibnamefont{Liu}},
  \bibinfo{author}{\bibfnamefont{Z.}~\bibnamefont{Zhang}},
  \bibinfo{author}{\bibfnamefont{X.}~\bibnamefont{Feng}},
  \bibinfo{author}{\bibfnamefont{K.}~\bibnamefont{Li}},
  \bibinfo{author}{\bibfnamefont{L.-L.} \bibnamefont{Wang}},
  \bibinfo{author}{\bibfnamefont{X.}~\bibnamefont{Chen}},
  \bibinfo{author}{\bibfnamefont{X.}~\bibnamefont{Dai}},
  \bibinfo{author}{\bibfnamefont{Z.}~\bibnamefont{Fang}}, \bibnamefont{et~al.},
  \bibinfo{journal}{Advanced materials} \textbf{\bibinfo{volume}{25}},
  \bibinfo{pages}{1065} (\bibinfo{year}{2013}{\natexlab{b}}).

\bibitem[{\citenamefont{Checkelsky et~al.}(2014)\citenamefont{Checkelsky,
  Yoshimi, Tsukazaki, Takahashi, Kozuka, Falson, Kawasaki, and
  Tokura}}]{checkelsky2014trajectory}
\bibinfo{author}{\bibfnamefont{J.~G.} \bibnamefont{Checkelsky}},
  \bibinfo{author}{\bibfnamefont{R.}~\bibnamefont{Yoshimi}},
  \bibinfo{author}{\bibfnamefont{A.}~\bibnamefont{Tsukazaki}},
  \bibinfo{author}{\bibfnamefont{K.~S.} \bibnamefont{Takahashi}},
  \bibinfo{author}{\bibfnamefont{Y.}~\bibnamefont{Kozuka}},
  \bibinfo{author}{\bibfnamefont{J.}~\bibnamefont{Falson}},
  \bibinfo{author}{\bibfnamefont{M.}~\bibnamefont{Kawasaki}}, \bibnamefont{and}
  \bibinfo{author}{\bibfnamefont{Y.}~\bibnamefont{Tokura}},
  \bibinfo{journal}{Nature Physics} \textbf{\bibinfo{volume}{10}},
  \bibinfo{pages}{731 EP } (\bibinfo{year}{2014}),
  \urlprefix\url{http://dx.doi.org/10.1038/nphys3053}.

\bibitem[{\citenamefont{Kou et~al.}(2014)\citenamefont{Kou, Guo, Fan, Pan,
  Lang, Jiang, Shao, Nie, Murata, Tang et~al.}}]{kou2014scale}
\bibinfo{author}{\bibfnamefont{X.}~\bibnamefont{Kou}},
  \bibinfo{author}{\bibfnamefont{S.-T.} \bibnamefont{Guo}},
  \bibinfo{author}{\bibfnamefont{Y.}~\bibnamefont{Fan}},
  \bibinfo{author}{\bibfnamefont{L.}~\bibnamefont{Pan}},
  \bibinfo{author}{\bibfnamefont{M.}~\bibnamefont{Lang}},
  \bibinfo{author}{\bibfnamefont{Y.}~\bibnamefont{Jiang}},
  \bibinfo{author}{\bibfnamefont{Q.}~\bibnamefont{Shao}},
  \bibinfo{author}{\bibfnamefont{T.}~\bibnamefont{Nie}},
  \bibinfo{author}{\bibfnamefont{K.}~\bibnamefont{Murata}},
  \bibinfo{author}{\bibfnamefont{J.}~\bibnamefont{Tang}}, \bibnamefont{et~al.},
  \bibinfo{journal}{Phys. Rev. Lett.} \textbf{\bibinfo{volume}{113}},
  \bibinfo{pages}{137201} (\bibinfo{year}{2014}),
  \urlprefix\url{https://link.aps.org/doi/10.1103/PhysRevLett.113.137201}.

\bibitem[{\citenamefont{Kurebayashi and Nomura}(2014)}]{kurebayashi2014weyl}
\bibinfo{author}{\bibfnamefont{D.}~\bibnamefont{Kurebayashi}} \bibnamefont{and}
  \bibinfo{author}{\bibfnamefont{K.}~\bibnamefont{Nomura}},
  \bibinfo{journal}{Journal of the Physical Society of Japan}
  \textbf{\bibinfo{volume}{83}}, \bibinfo{pages}{063709}
  (\bibinfo{year}{2014}).

\bibitem[{\citenamefont{Habe and Asano}(2014)}]{habe2014three}
\bibinfo{author}{\bibfnamefont{T.}~\bibnamefont{Habe}} \bibnamefont{and}
  \bibinfo{author}{\bibfnamefont{Y.}~\bibnamefont{Asano}},
  \bibinfo{journal}{Physical Review B} \textbf{\bibinfo{volume}{89}},
  \bibinfo{pages}{115203} (\bibinfo{year}{2014}).

\bibitem[{\citenamefont{Burkov}(2018)}]{burkov2018mirror}
\bibinfo{author}{\bibfnamefont{A.~A.} \bibnamefont{Burkov}},
  \bibinfo{journal}{Phys. Rev. Lett.} \textbf{\bibinfo{volume}{120}},
  \bibinfo{pages}{016603} (\bibinfo{year}{2018}).

\bibitem[{\citenamefont{Ueda et~al.}(2015)\citenamefont{Ueda, Fujioka, Yang,
  Shiogai, Tsukazaki, Nakamura, Awaji, Nagaosa, and Tokura}}]{ueda2015magnetic}
\bibinfo{author}{\bibfnamefont{K.}~\bibnamefont{Ueda}},
  \bibinfo{author}{\bibfnamefont{J.}~\bibnamefont{Fujioka}},
  \bibinfo{author}{\bibfnamefont{B.-J.} \bibnamefont{Yang}},
  \bibinfo{author}{\bibfnamefont{J.}~\bibnamefont{Shiogai}},
  \bibinfo{author}{\bibfnamefont{A.}~\bibnamefont{Tsukazaki}},
  \bibinfo{author}{\bibfnamefont{S.}~\bibnamefont{Nakamura}},
  \bibinfo{author}{\bibfnamefont{S.}~\bibnamefont{Awaji}},
  \bibinfo{author}{\bibfnamefont{N.}~\bibnamefont{Nagaosa}}, \bibnamefont{and}
  \bibinfo{author}{\bibfnamefont{Y.}~\bibnamefont{Tokura}},
  \bibinfo{journal}{Physical review letters} \textbf{\bibinfo{volume}{115}},
  \bibinfo{pages}{056402} (\bibinfo{year}{2015}).

\bibitem[{\citenamefont{Jin et~al.}(2017)\citenamefont{Jin, Wang, Chen, Zhao,
  Zhao, and Xu}}]{jin2017ferromagnetic}
\bibinfo{author}{\bibfnamefont{Y.~J.} \bibnamefont{Jin}},
  \bibinfo{author}{\bibfnamefont{R.}~\bibnamefont{Wang}},
  \bibinfo{author}{\bibfnamefont{Z.~J.} \bibnamefont{Chen}},
  \bibinfo{author}{\bibfnamefont{J.~Z.} \bibnamefont{Zhao}},
  \bibinfo{author}{\bibfnamefont{Y.~J.} \bibnamefont{Zhao}}, \bibnamefont{and}
  \bibinfo{author}{\bibfnamefont{H.}~\bibnamefont{Xu}}, \bibinfo{journal}{Phys.
  Rev. B} \textbf{\bibinfo{volume}{96}}, \bibinfo{pages}{201102}
  (\bibinfo{year}{2017}),
  \urlprefix\url{https://link.aps.org/doi/10.1103/PhysRevB.96.201102}.

\bibitem[{\citenamefont{Zhang et~al.}(2009)\citenamefont{Zhang, Liu, Qi, Dai,
  Fang, and Zhang}}]{zhang2009topological}
\bibinfo{author}{\bibfnamefont{H.}~\bibnamefont{Zhang}},
  \bibinfo{author}{\bibfnamefont{C.-X.} \bibnamefont{Liu}},
  \bibinfo{author}{\bibfnamefont{X.-L.} \bibnamefont{Qi}},
  \bibinfo{author}{\bibfnamefont{X.}~\bibnamefont{Dai}},
  \bibinfo{author}{\bibfnamefont{Z.}~\bibnamefont{Fang}}, \bibnamefont{and}
  \bibinfo{author}{\bibfnamefont{S.-C.} \bibnamefont{Zhang}},
  \bibinfo{journal}{Nature physics} \textbf{\bibinfo{volume}{5}},
  \bibinfo{pages}{438} (\bibinfo{year}{2009}).

\bibitem[{\citenamefont{Liu et~al.}(2010)\citenamefont{Liu, Qi, Zhang, Dai,
  Fang, and Zhang}}]{liu2010model}
\bibinfo{author}{\bibfnamefont{C.-X.} \bibnamefont{Liu}},
  \bibinfo{author}{\bibfnamefont{X.-L.} \bibnamefont{Qi}},
  \bibinfo{author}{\bibfnamefont{H.}~\bibnamefont{Zhang}},
  \bibinfo{author}{\bibfnamefont{X.}~\bibnamefont{Dai}},
  \bibinfo{author}{\bibfnamefont{Z.}~\bibnamefont{Fang}}, \bibnamefont{and}
  \bibinfo{author}{\bibfnamefont{S.-C.} \bibnamefont{Zhang}},
  \bibinfo{journal}{Physical Review B} \textbf{\bibinfo{volume}{82}},
  \bibinfo{pages}{045122} (\bibinfo{year}{2010}).

\bibitem[{\citenamefont{Wakatsuki et~al.}(2015)\citenamefont{Wakatsuki, Ezawa,
  and Nagaosa}}]{wakatsuki2015domain}
\bibinfo{author}{\bibfnamefont{R.}~\bibnamefont{Wakatsuki}},
  \bibinfo{author}{\bibfnamefont{M.}~\bibnamefont{Ezawa}}, \bibnamefont{and}
  \bibinfo{author}{\bibfnamefont{N.}~\bibnamefont{Nagaosa}},
  \bibinfo{journal}{Scientific Reports} \textbf{\bibinfo{volume}{5}},
  \bibinfo{pages}{13638 EP } (\bibinfo{year}{2015}), \bibinfo{note}{article},
  \urlprefix\url{http://dx.doi.org/10.1038/srep13638}.

\end{thebibliography}

\end{document}